\documentclass[referee]{aa} % for a referee version
\usepackage{graphicx}
\usepackage{txfonts}
\begin{document}

   \title{The dipper light curve of V715 Per: is there dust in the magnetosphere?}

%   \subtitle{Is there dust in the magnetosphere\?}

   \author{E. Nagel
          \inst{1}
          \and
          J. Bouvier\inst{2}
          }

   \institute{Departamento de Astronom\'ia, Universidad de Guanajuato,
	       Callej\'on de Jalisco S/N,Guanajuato,Gto,36240,M\'exico
              \email{e.nagel@ugto.mx}
         \and
         Univ. Grenoble Alpes, CNRS, IPAG, 38000 Grenoble, France\\
              \email{jerome.bouvier@obs.ujf-grenoble.fr}
             }

   \date{Received ...., 2020; accepted .....}

% \abstract{}{}{}{}{} 
% 5 {} token are mandatory
 
  \abstract
   {The dipper optical light curves in young stellar objects are commonly interpreted as
partial or total occultation of the stellar radiation by dust surrounding the star.}
   {In this work, we analyze the amplitude of the optical light curve of V715 Per, located in the young star forming region IC 348. Observations gathered over the years suggest that the light curve can be explained by dust extinction events.}
   {In our model, the dust is distributed inside the magnetosphere according to the strength of the stellar
magnetic field. The dust distribution is modulated by the vertical component of the field, whose axis is misaligned with
respect to the rotational axis. We include a model for the evaporation of the dust reaching the magnetosphere in order to consistently calculate its distribution.}
   {For V715 Per, there is dust in the optically thick warp at the disk truncation radius. We suggest that 
the optical light curve is explained by extinction caused by dust reaching inside the 
magnetosphere. The dust distribution is optically thin and due to the high 
temperature and low density, it cannot survive for a long time. However because
the grains rapidly move towards the stellar surface and the sublimation is 
not instantaneous, there is a layer of dust covering the magnetosphere responsible
for the extinction.}
   {Dust surviving the harsh conditions of the magnetospheric accretion flow may be responsible for some of the dipper light curves.}

   \keywords{Accretion, accretion disks --
                circumstellar matter 
               }

   \maketitle
%
%-------------------------------------------------------------------

\section{Introduction}

The study of optical and near-infrared light curves of young stellar objects (YSOs) provides information about the dust distribution around the object. Observed light curves of YSOs clearly
show its variability \citep{alencar,cody,morales,stauffer2015,stauffer2016}, the different features allow to define different classes, suggesting
the rich and complex dynamics ocurring in the dusty disk. For the modeling of light curves 
showing periodicity of a few days, it is required to take into account the interaction of the
magnetic field lines defining the stellar magnetosphere with the inner edge of the disk, in particular for dipper light curves. The 
magneto-hydrodynamical simulations by \citet{romanova} show the formation of 
magnetospheric streams and a bending wave in the innermost disk regions. These time-dependent
structures with variable height can be used to interpret the light curves.

For a YSO, the optical flux mainly comes from the star, thus, its variability either comes from changes of the spot distribution at the stellar surface or results from the extinction of the stellar light by circumstellar dust. We are analyzing the second mechanism here, such that 
we can directly extract the characteristics of the azimuthal dust distribution from the shape of the light curve. If the 
structures blocking the stellar radiation are optically thick, the color calculated between two
optical wavelengths is constant in terms of phase. An object which has shown this kind of behavior in the past is AA Tau whose light curve can be interpreted with a non-axisymmetric optically thick warp in the
innermost regions of the disk surrounding the star \citep{bouvier}. They suggest that
the warp is formed at the base of the magnetospheric streams caused by an inclined stellar
magnetic field interacting with the disk. This idea was followed by others: 
\citet{fonseca} studied V354 Mon, \citet{mcginnis} applied this for a large sample in NGC 2264, and \citet{nagel2019} studied the objects Mon-660, Mon-811, Mon-1140 and
Mon-1308. The optical light curves analyzed in these works come from the CoRoT Space Telescope.
Simultaneous observations with the Spitzer Space Telescope in the infrared allows to connect
the behavior in both wavelength ranges, interpreted via the dust distributed around the star
\citep{mcginnis,nagel2019}.

The YSO V715 Per is one of several objects that present light curves showing color changes consistent with circumstellar extinction \citep{barsunova2013,barsunova2015,barsunova2016}. This behaviour
is common in young UX Ori stars, one of the main characteristics of this type is stochastic
brightness variations caused by extinction where the dust is located in the circumstellar
environment. \citet{grinin2018} analyze V715 Per in the vein of their previous works, where for 5 objects coming from a large set of T Tauri stars, the light curves can be explained by extinction \citep{barsunova2013,barsunova2015,barsunova2016}. Also \citet{sicilia} show that
the quasi-periodic eclipses in the young dipper RX J1604.3-2130A can be explained with extinction at the base of the magnetospheric streams.

The stellar parameters of V715 Per are listed in Table~\ref{table:stellar_parameters}. The first value for the stellar mass ($M_{\star}$) comes
from \citet{kirk} and the one in parenthesis comes from \citet{leblanc}. The values are different because \citet{leblanc} use the pre-main-sequence evolutionary models calculated by \citet{dantona} while \citet{kirk} use the evolutionary tracks by 
\citet{baraffe}. The difference also depends on the way the bolometric luminosity $L_{bol}$ is calculated because besides the spectral type, \citet{leblanc}'s value uses the J-band magnitude and \citet{kirk}'s value uses the estimated age of the object (1Myr). The stellar radius ($R_{\star}$) is from \citet{leblanc} and the stellar effective temperature ($T_{\star}$) is taken from \citet{luhman} using the associated spectral type (K6) shown in 
\citet{herbig}. The mass accretion rate ($\dot{M}$) is taken from \citet{dahm} where we take the mean of the values associated to the estimates using the $HeI$, 
$H\alpha$ and $CaII$ lines. We remove the estimate using the R-band veiling because as 
mentioned in \citet{dodin}, if the veiling by lines is included, the depth changes
varies for different photospheric lines such that the estimate of $\dot{M}$ is not
reliable in this case. \citet{ruiz} observed the system with ALMA but the result was a non-detection, preventing them from deriving an inclination for the system. 

The aim of this paper is to present a model to explain the amplitude and period
of the optical light curve of V715 Per showed in \citet{grinin2018}. In our model, the variation of the magnitude comes from phase-dependent extinction due to material located in the magnetospheric streams. 

In Section~\ref{sec-charact} we describe the characteristics of the optical
light curve of V715 Per, in Section~\ref{sec-model} we show the relevant aspects
of the modeling, in Section~\ref{sec-dust} we address two important questions, namely whether 
there is dust in the disk and whether there is dust in the magnetosphere. The free
and fixed parameters of the model are described in Section~\ref{sec-parameters} and in 
Section~\ref{sec-modeling}, the model is applied to V715 Per. Finally in
Section~\ref{sec-conclusions} we present the conclusions of this study.

\begin{table}
\caption{Parameters for V715 Per}             
\label{table:stellar_parameters}      
\centering                          
\begin{tabular}{c c c c}        
\hline\hline                 
$M_{\star}(M_{\odot})$ & $R_{\star}(R_{\odot})$ & $T_{\star}(K)$ & 
$\dot{M}(M_{\odot}yr^{-1})$ \\     
\hline                        
 0.906(0.56) & 2.31 & 4205 & $5\times 10^{-9}$  \\      
\hline                                   
\end{tabular}
\end{table}

%--------------------------------------------------------------------
\section{The characteristics of the periodic light curve of V715 Per}
\label{sec-charact}

\citet{grinin2018} present observations for the stellar system V715 Per in
the optical (VRI bands) and in the infrared (JHK bands). They present photometric
data compiled from different campaigns between 2003 and 2017. For this large
time scale, they confirm the presence of low-amplitude periodic brightness 
variations $\Delta mag\sim 0.1$ with a period equal to 5.25days. 
\citet{flaherty} 
photometrically studied this system with the Spitzer Space Telescope at $3.6$
and $4.5\mu$m over 40 days and detected a signal with a period of 14.7days, 
which is not the stellar rotation period.
The keplerian radius corresponding to 5.25days is $0.0574$AU. The bending wave formed by the hydrodynamical interaction of the stellar magnetic dipolar field and the gaseous disk \citep{romanova} is used for the physical interpretation of 
the data presented in \citet{grinin2018}. Because of the larger period, the IR light curve cannot be analyzed with the model used here, thus it is required another
mechanism to explain the photometric Spitzer data \citep{flaherty} for this system.
The aim of this work is explain the amplitude of the optical light curve. 
The interpretation of its details requires full 3D-MHD simulations which take into
account each scenario in a highly dynamical region. This endeavor is well beyond the goal of this work.

In \citet{grinin2018}, the observed period in the light curve of 
$P=5.25$days is interpreted by occultations of material located at the 
keplerian radius consistent with $P$. At this location, they assume that
the material accreting radially in the disk is halted by the magnetic
pressure associated with the stellar magnetic field (hereafter referred to as B). In other words, this
radius also corresponds to the magnetospheric radius ($R_{mag}$) such
that the occulting material belongs to the base of the magnetospheric streams, near the plane of the disk that 
transport material towards the stellar surface at the magnetic poles, 
where two hot spots are formed. 

\section{The model used to interpret the observations}
\label{sec-model}

\subsection{The system geometry and physical mechanisms}
\label{sec-system}

The 5.25day period is the one we are trying to interpret. A model consistent with this period involves material close to the star.
The dust responsible to produce the stellar eclipses accretes in the disk 
reaching the outskirts of the magnetosphere located at the radius 
$R_{mag}$. The existence of dust inside the magnetosphere requires that the sublimation radius ($R_{sub}$) is smaller than $R_{mag}$. The disk regions close to the midplane are optically thick to the
stellar radiation, thus, they are responsible to produce total or partial 
stellar eclipses. For the well-known system AA-Tau, the photometric data 
shaping the periodical optical light curve do not show color changes 
\citep{bouvier}, thus,
the dust responsible for it should be optically thick, which is in accord 
to an optically thick disk. For V715 Per the periodic brightness variations
show color changes consistent with circumstellar extinction, thus this material
should be located in an optically thin environment. For the model presented 
here, this region corresponds to the dust located in the magnetospheric 
streams. Note that the radial dust distribution is
not important for this estimate because the matter is added along the line of
sight (from now on los). 

We use the model for the accretion along a non-inclined dipolar B presented 
in \citet{hartmann} where the density is given by

\begin{equation}
 \rho_{H}(R,\theta)={\dot{M}R_{\star}\times R^{-5/2}\times (4-3sen^2\theta)^{1/2} \over 4\pi({R_{\star}\over R_{in}}-{R_{\star}\over R_{out}})
       (2GM_{\star})^{1/2}\times (1-sen^2\theta)^{1/2}}.
\end{equation}

Here, $G$ is the gravitational constant, $R$ is the spherical radius, $\theta$ is 
the poloidal angle,  
$R_{in}$ and $R_{out}$ corresponds to the inner and outer intersections
of the magnetospheric stream with the disk plane. We fix these values as
$R_{in}=0.9R_{mag}$ and $R_{out}=R_{mag}$, as an order of magnitude estimate
motivated by the simulations presented in Figure 8 in \citet{romanova}.
When it is used in the code, we take into account that the dipole is inclined with
respect to the rotational axis. As in \citet{hartmann}, the material velocity
corresponds to ballistic infall from rest following the B lines. Finally, the 
gas density inside the magnetosphere ($\rho_{mag}$) is given by

\begin{equation}
 \rho_{mag}(R,\theta,\phi)=\rho_{H}(R,\theta)*\gamma*{B_{pol}(\theta,\phi)\over 
B_{dip}},
\label{eq:rho_mag}
\end{equation}

where $\phi$ is the azimuthal angle, $B_{pol}$ is the poloidal magnetic field and $B$ at $R_{\star}$ is given by $B_{dip}$. This factor is included in 
order to modulate the density such that the gas is concentrated where the
magnetic pressure is larger. The value for the free-adimensional parameter 
$\gamma$ is such that $\rho_{mag}$ is consistent with $\dot{M}$. Because the magnetic axis is inclined with respect to the rotational axis, $B_{pol}$ depends on $\phi$ and $\theta$. All this together, the 
gas distribution depends on the azimuthal ($\phi$) and poloidal angles 
($\theta$) such that the optical depth varies for different los. The presence and amount of dust is estimated using the temperature calculated as 
an equilibrium between the heating of stellar radiation and cooling as a 
blackbody at such temperature. 
Two facts guarantee the presence of 
dust in the magnetosphere: 1) the grains do not sublimate instantaneously
and 2) the free-fall time ($t_{ff}$) is just a few hours. This means that 
the dust can survive in a layer with an outer radius given by $R_{mag}$, 
its inner boundary depends on the timescale of sublimation ($t_{sub}$) 
and $t_{ff}$. In our modeling, the amount of dust in this layer is 
responsible to explain the deepness of the light curve at the various 
phases, as described in Section~\ref{sec-modeling}.

\subsection{Details of the code used}

According to a dust distribution around the star, the code calculates
the total or partial extinction of the stellar radiation heading to
the observer. For our particular aim, the dust is located inside the
magnetosphere, which is limited by $R_{mag}$. The dust reservoir is 
the protoplanetary disk where the grains are slowly moving towards 
the star. At $R_{mag}$, the material find a radial equilibrium
because the magnetic pressure is equal to the ram pressure, such that
the vertical direction is the only available path to follow. As we
are assuming that the gas ionization is enough for the particles 
to be frozen to the $B$-lines, then the gas and the dust attached 
will follow the path given by the $B$-lines almost perpendicularly 
to the midplane of the disk. The material leaving the disk form
the magnetospheric streams which consist of $B$-lines that shrinks as 
$B$ is increased when it approaches the stellar surface. This behavior 
is taken into account to calculate the expression for $\rho_{mag}$ 
(see equation~\ref{eq:rho_mag}). 

We assume that $B$ is given by a dipolar 
magnetic field which is tilted with respect to the rotational axis by
an angle $\beta_{dip}$ \citep{gregory} with $B$ at $R_{\star}$ 
given by $B_{dip}$. In order to find $B_{dip}$, we use 
eq. 1 in \citet{konigl} substituting the 
values for $R_{mag}$, $M_{\star}$ and $\dot{M}$. For the model presented here, 
$\beta_{dip}=5deg$ and $B_{dip}=398G$. Along with $\beta_{dip}$ and 
$B_{dip}$, the longitude where the magnetic axis is tilted 
($\psi_{dip}$) is the other free parameter that defines the stellar
$B$. Because the star is azimuthally rotating, every section of the
stellar surface crosses the los once every rotating period, which 
according to the modeling is equal to $P$. This means that the value
$\psi_{dip}$ is only relevant to locate the magnetic structure
with respect to the light curve phases without changing the shape 
of the latter.

\section{Presence of dust}
\label{sec-dust}

\subsection{Is there dust in the disk at $R_{mag}$?}
\label{sec-dust-disk}

For the stellar radiation that do not show color changes as in 
AA-Tau \citep{bouvier}, the occulting structure should be
optically thick. For the stellar system modeled here, this 
corresponds to the optically thick disk that is perturbed in the
vertical direction to form the required structures according 
to each case. In this section, we answer the question, is there
optically thick dust at $R_{mag}$ for the stellar system V715 Per?

We use the stellar and disk parameters associated to this object \citep{grinin2018}
which are shown in Table~\ref{table:stellar_parameters} to calculate the disk 
surface density 

\begin{equation}
 \Sigma={\dot{M}\over 3\pi\alpha}{\Omega\mu m_{H}\over KT_{c}}[1-\sqrt{R_{\star}/R}]
\end{equation}

given by \citet{lyndenbell} for a viscous stationary disk.  
The viscosity is parameterized with $\alpha$ (we assume $\alpha=0.01$ as typical),
$\Omega$ is the keplerian angular velocity, $\mu$ is the molecular number,
$m_{H}$ is the hydrogen mass, $K$ is the Boltzmann constant, $T_{c}$ is the
temperature at the midplane and $R$ is the distance towards the star.

Using $\Sigma$, we calculate the disk density at $R_{mag}$ with 

\begin{equation}
 \rho_{disk}={\Sigma\over 2H},
\end{equation}

where $H=c_{s}/\Omega$ is the hydrostatic scale height and 
$c_{s}=\sqrt{KT_{c}\over \mu m_{H}}$ is the sound speed. This results in

\begin{equation}
 \rho_{disk}(R=R_{mag})=7.59*10^{-9}\times({\dot{M}\over 5\times 10^{-9}
                         M_{\odot}yr^{-1}})\times(1000/T_{c})gcm^{-3}.
\end{equation}

Using a linear interpolation of the Table 3 of \citet{pollack}
showing the sublimation temperature ($T_{sub}$) in terms of $\rho_{disk}$,
we estimate $T_{sub}=1417K$ at $R=R_{mag}$. We use 
$\dot{M}=5\times 10^{-9}M_{\odot}yr^{-1}$ and $T_{c}=1000K$ as typical 
for the disk.

The temperature in the warp ($T_{mag}=T_{c}(R_{mag})$) is calculated 
using the model 
of an optically thick wall heated by the stellar radiation presented in 
\citet{dalessio} and used in \citet{nagel2010} to calculate the 
emission of the wall heated by a binary stellar system. For the stellar 
parameters given in Table~\ref{table:stellar_parameters} and a typical 
dust composition and the grain size distribution proportional to the power-law 
$s^{-3.5}$ between a minimum grain size $s_{min}=0.005\mu$m (according 
to the value used to characterize the interstellar medium (ISM), Mathis et al. \citeyear{mathis}) and a maximum grain size 
$s_{max}=1mm$ (assuming that in the disk the grains increase their size
from $s_{max}=0.25$ typical of the ISM), $T_{mag}=1309K$. 
Because $T_{mag}\le T_{sub}$ then the answer to the stated question is: 
the warp contains dust, such that the assumption of an optically thick 
structure is valid.
Note that for V715 Per, the affirmative answer is important because 
this means that the disk at $R_{mag}$ is the dust reservoir for the 
magnetospheric streams.     

\subsection{Is there dust inside the magnetosphere?}
\label{sec-dust-mag}

The disk material at $R_{mag}$ is forced to move along the B-lines 
(because the magnetic pressure is equal to the ram pressure) filling
the magnetosphere with a specific gas distribution. The density in 
this region allow us to calculate $T_{sub}$ and with a comparison with
the dust temperature $T_{dust}$ we can check if there is dust in the
magnetosphere as it is required to explain the light curve of V715 Per.
In order to check if there is dust in the magnetosphere, we follow 
the next simple estimate. The use of $\rho_{mag}$ from equation~\ref{eq:rho_mag} is included in the full modeling. The gas density in this region $\rho_{mag}$ depends on $\dot{M}$ and the time it takes to arrive to the stellar surface starting its journey at $R_{mag}$. The mean of $\rho_{mag}$ is calculated as

\begin{equation}
 <\rho_{mag}>=M_{gas,mag}/V_{mag},
\end{equation}

where $V_{mag}={4\pi\over 3} R_{mag}^3$ is the volume of the magnetosphere
and $M_{gas,mag}$ is the gaseous mass inside the magnetosphere, estimated
with

\begin{equation}
 M_{gas,mass}=\dot{M}\times t_{ff},
\end{equation}

where $t_{ff}=R_{mag}/v_{ff}$ with the free-fall velocity given by

\begin{equation}
 v_{ff}=\sqrt{2GM_{\star}/R_{\star}}\sqrt{1-R_{\star}/R_{mag}}
\end{equation}

Doing the required substitution in the previous equations, 
we find that $<\rho_{mag}>=2.93(5.29)\times 10^{-15}g\,cm^{-3}$ and
$T_{sub}=1019(1038)$K for $M_{\star}=0.906(0.56)M_{\odot}$. The two 
values for $T_{sub}$ correspond to the two values for $M_{\star}$ (see Table~\ref{table:stellar_parameters}). It is important to note that both values for $T_{sub}$ 
are similar such that the uncertainty in $M_{\star}$ is not relevant for the 
results presented here. Assuming a more realistic estimate for $<\rho_{mag}>$,
considering that the material is restricted inside an axisymmetric stream limited by magnetic field lines in the dipolar configuration ($r\propto sin^{2}\theta$), we 
estimate that only $12\%$ of the magnetospheric volume is covered with material. 
Because $T_{sub}$ is not highly dependent on $\rho$, the new estimate ($T_{sub}=1094(1106)$K) is close to the previous one. The estimate
of $T_{sub}$ comes from a linear interpolation of Table 3 of \citet{pollack}. Assuming an optically thin environment with typical
dust composition and grains size distribution (see 
Section~\ref{sec-dust-disk}), the estimated temperature in the 
magnetosphere is $T_{mag}\sim 1309$K which is larger than $T_{sub}$. 
This temperature is at the outskirts of the magnetosphere. There is not dust
in a region close to the magnetospheric spots, where the temperature is up to 
$\sim 7000$K. Starting at some minimum radius, there is a shell of evaporating dust
up to $R_{mag}$. In order to find the gas temperature inside the magnetosphere, \citet{hartmann} assume a volumetric heating rate 
$\propto R^{-3}$ coming from a flux of Alfven waves during the gas accretion,
which is cooled with a given radiative cooling rate.
We conclude from this estimate that the grains will evaporate inside
the magnetosphere.

In a study of the formation and evolution of chondrules inside a 
protoplanetary disk, \citet{shu} estimate that the
sublimation timescale, $t_{sub}$, is about a
few hours when the grains are heated to above $T_{sub}$.
Using the stellar parameters from Table~\ref{table:stellar_parameters}, 
$t_{ff}=6.79(7.56)$hours ($M_{\star}=0.906(0.56)M_{\odot}$). 
Thus it is reasonable to assume the existence of a dusty layer in the
funnel flow because $t_{sub}\sim t_{ff}$.

For an estimate of the width of this layer, we included in this work 
the evaporating process as is described in \citet{xu}. The rate 
at which the grain size is reduced at the temperature $T_{dust}$ is given by 

\begin{equation}
 \dot{s}=J(T_{dust})/\rho_{dust},
\label{eq:sdot}
\end{equation}

where $J$ is the net sublimation mass-loss flux and $\rho_{dust}$ is the
density of the dust grain, respectively. The expression
for $J$ is given by

\begin{equation}
 J(T_{dust})={\alpha_{subl}[P_{sat}-P_{g}]\over c_{s}\sqrt(2\pi)},
\end{equation}

and comes from the kinetic theory of gases. Here, $\alpha_{subl}$ is the
evaporation coefficient, $P_{sat}$ is the saturation pressure (where the
condensation and sublimation are in equilibrium), $P_{g}=\rho_{mag}\times c_{s}^2$
is the gas pressure and finally $c_{s}$ is the sound speed. Besides 
$P_{g}$, there are the gas ram pressure and the magnetic pressure which are
in equilibrium. For the modeling,
it is taken $\alpha_{subl}=0.1$ and $\rho_{dust}=3g\,cm^{-3}$ as typical values.
The expression for $P_{sat}$ is written as

\begin{equation}
 P_{sat}(T_{dust})=e^{-A/T_{dust}+B},
\end{equation}

with $A=65000K$ and $B=35$ for a generic refractory material \citep{xu}.
For the analysis presented in this work, $A$, $B$, $\rho_{dust}$ and 
$\alpha_{subl}$ are fixed but the specific characteristics of the system that
we do not know a priori will lead to changes in these values. 
The evaporation process occurs when $P_{sat}>P_{g}$ and condensation 
occurs when $P_{sat}<P_{g}$. For example at $T_{dust}=1500$K (typical value for dust sublimation), the value for $P_{sat}$ leads to condensation instead of sublimation. For $T_{dust}=2000$K,
$P_{sat}$ increases in 5 orders of magnitude, such that the evaporation
stage is reached. For $T_{dust}=7000$K, $P_{sat}$ increases 10 orders of 
magnitude more such that the dust only survives in a layer close to $R_{mag}$
where $T_{dust}$ is close to $1500$K.

As mentioned in Section~\ref{sec-dust-disk}
the grains are distributed according to a power-law, in the Appendix~\ref{sec-appendix} we  
show that during the evaporation process the power of the size distribution is 
conserved. Also in the Appendix~\ref{sec-appendix}, we explain how the opacity at each point
in a stationary stream is calculated which includes the fact that the evaporation
decreases the dust abundance. 

Because the power-law shape of the size distribution is conserved, the
survival of dust can be analyzed with the survival of the grain in the 
magnetosphere with the largest size $s_{max}$ 
using equation~\ref{eq:sdot}. We describe this process using a two steps 
algorithm. The steps are: 
1) For each point in the magnetosphere we identify if it is located over a
B-line that belongs to the stream limited by $R_{in}$ and $R_{out}$ (see 
Section~\ref{sec-system}). 2) Check if there is dust. The condition is:
if $s_{max}\geq 0.25\mu$m then there is dust 
at this point, see the Appendix~\ref{sec-appendix} for details. The grains 
with sizes smaller 
than $s=0.25\mu$m will quickly disappear in the harsh environment the 
particle face from there on, following its trajectory towards the 
stellar surface.    
We use the ballistic velocity field of \citet{hartmann}, the shape of the 
B-line and equation~\ref{eq:sdot} to describe 
the evaporation process along the trajectory. In this equation, we substitute
$\rho_{mag}$ from equation~\ref{eq:rho_mag}.

Repeating the two steps dust survival algorithm for every point in the
region surrounding $R_{mag}$ will lead to a dust 
distribution that depends on $R$, $\theta$ and $\phi$.
However, in order to estimate a typical radial
width of the dusty layer, we apply the algorithm along different los in order
to sample the complete region and
answer the question: How far in the stream the optically thin dust can survive?
This depends on its temperature ($T_{thin}$) which for a given dust composition,
grain size distribution and for a fixed stellar luminosity only depends on 
the distance to the star. For V715 Per at $R=0.4R_{mag}$, $T_{thin}=2221$K 
which corresponds to $P_{sat}>>P_{g}$. At this state, a grain of $s=1$mm 
sublimates in about $1$hr which is a factor of $7$ lower than $t_{ff}$,
which means that at this location a grain rapidly disappear and dust closer
to this radius is not present. At the outskirts of the stream at $R=R_{mag}$,
$T_{thin}=1479$K, $P_{sat}<P_{g}$ and the process is condensation instead of
sublimation. This clearly means that the dust incorporating to the stream 
survive and steadily decrease in size until is completely evaporated as
it moves in the stream towards the star. Between $R=0.7$ and 
$R=0.8R_{mag}$, $P_{sat}=P_{g}$ and the condensation turns into 
sublimation. This means that the dusty radial width is larger than 
$0.2R_{mag}$, thus, there is dust inside the magnetosphere that can shape
the optical lightcurve. This estimate is done using a distribution
of grains with $s_{max}=1$mm. In Section~\ref{sec-modeling}, we show that
according to the maximum grain size at the base of the stream, the
radial width of the dusty layer varies.

Note that when the gas transit from the disk to the stream, probably 
there are a selection on the size of the grains that will do the journey 
because the first grains that incorporates to the stream are the ones 
located in the disk atmosphere where its size distribution is assumed 
to be consistent with the dust in the ISM (this is because detailed disk
models, such as in \citet{dalessio2006}, show grain settling towards the 
midplane). Because we do not know
the details of this process, a reasonable assumption is to consider
that the grains size at the base of the stream are distributed 
according to a power-law with given $s_{max}$, which becomes the relevant
free parameter to tune the modeling. In the 
Appendix~\ref{sec-appendix}, details of the survival of dust as the 
material moves towards the star are shown.

\section{Parameters for the modeling}
\label{sec-parameters}

\subsection{Fixed parameters}
\label{sec-fixed} 

From \citet{grinin2018}, we extract the stellar parameters 
($M_{\star}$,
$R_{\star}$,$T_{\star}$) noting that there are two values for $M_{\star}$:
$0.56M_{\odot}$ \citep{leblanc} and $0.906M_{\odot}$ 
\citep{kirk}. The distance to IC 348, where the source is located
is $315$pc. Using a mean of the estimates in \citet{dahm} 
we fix $\dot{M}=5\times 10^{-9}$. The light curve period is 
$P=5.25$days \citep{grinin2018}, value that is very close to the 
rotational period of the star, $P_{rot}=5.221$days \citep{cohen}.
The inclination of the system is assumed as $i=77\deg$, as the largest
value expected such that the outside part of the disk is not blocking the
stellar radiation \citep{nagel2019}. 

Variations in the size distribution of the dust grains will change
the opacity ($\kappa$) such that a consistent change on
the stream dust to gas mass fraction ($\zeta_{stream}$ ) will be 
required to explain the observations. In other words, there is
a degeneracy between the last two variables. However,
in the evaporation process that we include in the 
modeling, the grain size distribution is calculated
at each point in the stream. The standard dust/gas mass 
fractions are fixed at typical values, $\zeta_{sil}=0.004$ 
and $\zeta_{grap}=0.0025$ for the silicate and the
graphite components, such that $\zeta_{std}=\zeta_{sil}+\zeta_{grap}=
0.0065$. Using this, we fix $\kappa$.
In the same process $\zeta_{stream}$ is
calculated because surface material continuously
evaporates from each grain, as the particles moves from
the base of the stream towards the star.
This means that the spatial distribution
of dust including the grain size distribution (a power-law
with a consistently calculated $s_{max}$) is taken into 
account when using the algorithm for dust survival 
(see Section~\ref{sec-dust-mag}).

\subsection{Free parameters}
\label{sec-free}

According to the modeling, the dust responsible to shape the 
light curves is located in the magnetosphere. As $\rho_{mag}$ is 
consistent with the observed $\dot{M}$ then in order to explain 
the depth of the section of the light curve interpreted as the stellar
eclipse with the magnetospheric streams, we focus on one free 
parameter, the maximum grain size at the base of the stream 
$s_{max,base}$.

As it is explained in the Appendix~\ref{sec-appendix} the 
grain size distribution at the base of the
stream is a power-law with an exponent $-3.5$ with a minimum 
grain size $s_{min}=0.005\mu$m and the maximum grain size 
$s_{max,base}$ is taken as the only free parameter. Its range
is between $0.25\mu$m, the value associated to the interstellar
medium (ISM) \citep{mathis} and
$1$mm, a value typical of the densest environment in a
protoplanetary disk. Note that $\kappa$ changes more than
an order of magnitude between the large grains distribution
and the small grains distribution, thus, it can be tuned through 
$s_{max,base}$ to explain the amplitude of the lightcurve. 
Using $\kappa$, the gas density ($\rho_{g}$) and $\zeta_{stream}$ 
we can check on each
los that the dust is optically thin. Including the grain 
evaporation, the system state moves towards an optically 
thin scenario, the basic requirement to explain the observations.

\section{Modeling}
\label{sec-modeling}

The three states of the light curve are: 1) a 
$5.25$d periodic signal of $\Delta mag\sim 0.1$, 2) a $5.25$d 
periodic signal that decreases in amplitude ($\Delta mag < 0.1$) 
and 3) very deep Algol-like minima reaching amplitudes up to 
$\Delta mag = 1$ in a timescale of days. At any time of the 
observations there is a periodic signal with evolving amplitude,
the Algol-like minima is superposed to the periodic signal. 
However, in the latter case, the dominant behavior is the one 
with the larger amplitude signal. As mentioned by \citet{grinin2018}, 
all states are interpreted by stellar extinction, such that the 
changes on the amount of dust surrounding the object should explain the 
observed variability. The state 1 is the one we explain in detail in
Section~\ref{sec-low-state}, where we use dust distributed along
the B-lines consistent with the value used for $\dot{M}$ such
that we name it, the low-$\dot{M}$ state. 

The state 2 is interpreted by \citet{grinin2018}
based on the assumption that a larger $\dot{M}$ reduces $R_{mag}$ such 
that the blocking structure decreases its area resulting in
a decrease of $\Delta mag$. The variability of $\dot{M}$ is suggested by 
the measurements of $EW(H_{\alpha})$ at different times by \citet{herbig}, 
\citet{luhman1998} and \citet{dahm}. The first two were taken 
outside the epoch of enhanced activity represented in the observations by
the last one; thus it is sampled both regimes. Our interpretation differs from 
the previously stated: a larger $\dot{M}$ (leading to a more powerful
accretion shock) means a decrease on the amount of dust occulting the star 
because the stellar heating increases.
A state with a larger $\dot{M}$ is expected in Clasical T-Tauri stars 
caused by their intrinsic variability.
The Algol-like minima identifying the state 3 is simply an 
UX Ori-like event (stochastic fluctuations of the circumstellar extinction)
where material continuously accumulating is explosively released to the 
star. The duration of the Algol-like minima is about a week such
that the delivery of an unusual amount of gas (and dust) occurs 
continuously during (1-2) stellar rotations.

\subsection{The low-$\dot{M}$ state: fiducial models}
\label{sec-low-state}

The value for $\dot{M}$ for V715 Per is estimated by \citet{dahm}
using different optical and infrared lines. We name the models
presented in this section, the fiducial models because we are just including
the stellar heating as the only source of heating. 

All the interpretation of the results is based on two competing mechanisms, the first one is that when in all the cells of the system 
$P_{sat}<P_{g}$ the system is on the non-evaporating stage, on the other
hand when $P_{sat}>P_{g}$ the system is in the evaporating stage. Note that
an increment of $s_{max,base}$ decrease $T_{dust}$ and with this 
$P_{sat}$ decreases, eventually making the system prone to condensation instead of sublimation, resulting in a large optical depth. Decreasing $s_{max,base}$
from its maximum value, the system changes from the non-evaporating stage
to the evaporating stage. The latter stage is mixed with the other mechanism:
an increment of the grain size decreases $\kappa$, resulting in a decrement of the optical depth ($\tau$). Thus, in the evaporating stage, the evaporation
in itself reduces the opacity because an amount of dust disappears but on the
other hand the evaporation reduces the grain size and the opacity increases.

\begin{figure} 
   \centering
   \includegraphics[angle=-90,width=8cm]{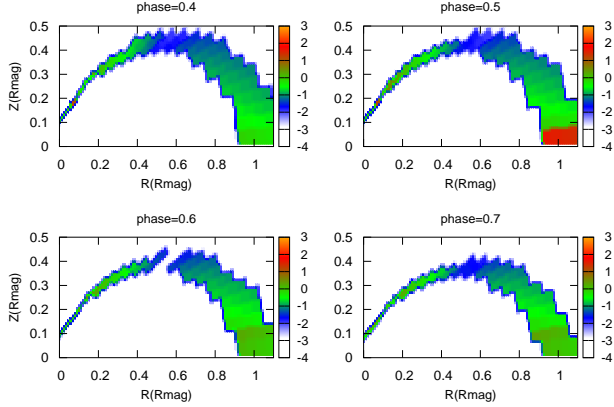}
      \caption{The color map representing the density for different phases. 
               This is part of the data used to find the light curves in 
               Figure~\ref{fig:lc}. Here, we show the model with 
               $s_{max,base}=1$mm.
              }
         \label{fig:dens}
   \end{figure}

\begin{figure} 
   \centering
   \includegraphics[angle=-90,width=8cm]{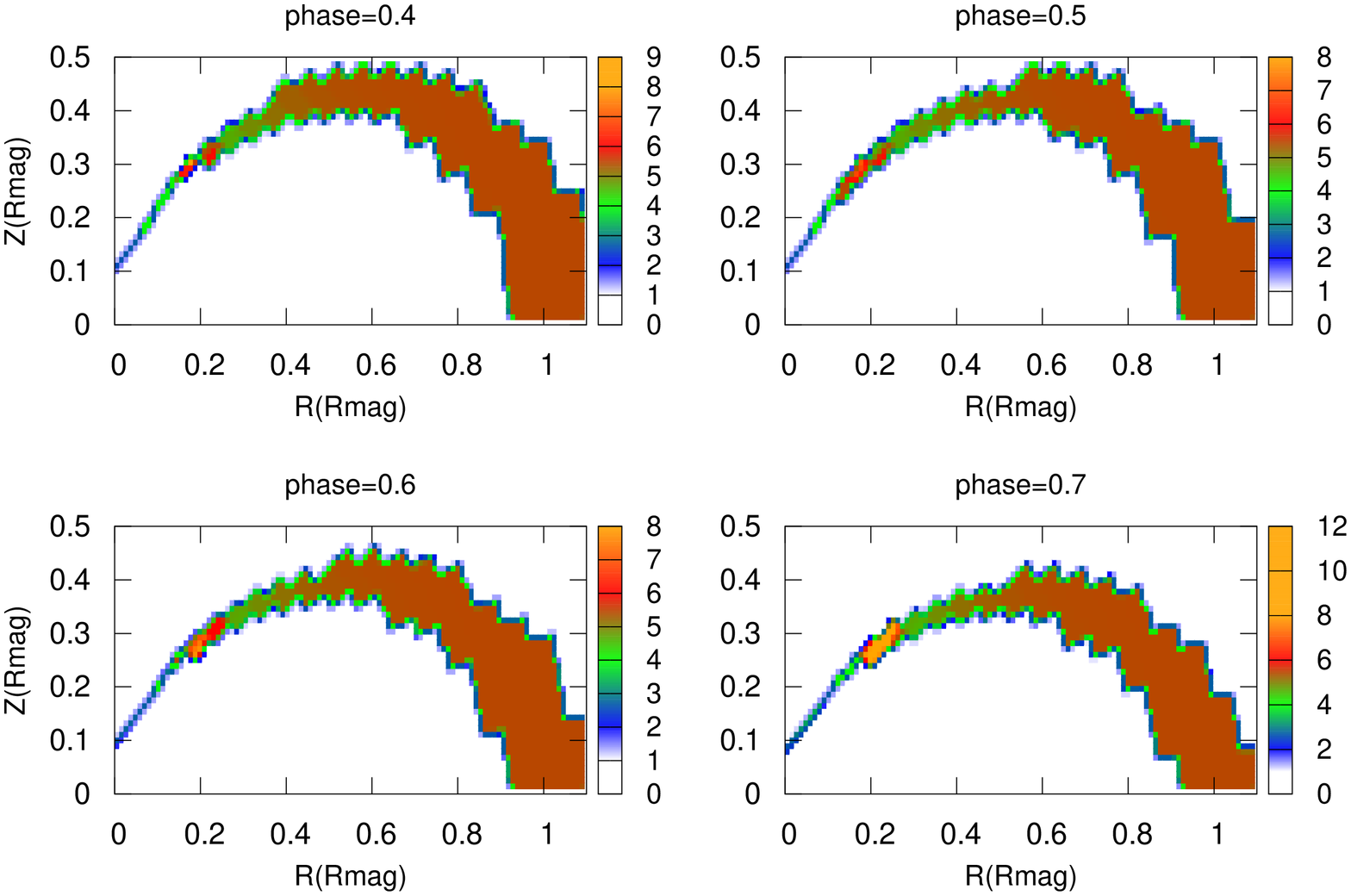}
      \caption{The color map representing the opacity for different phases.  
               This is part of the data used to find the light curves in 
               Figure~\ref{fig:lc}. Here, we show the model with 
               $s_{max,base}=1$mm. 
              }
         \label{fig:opa}
   \end{figure}

The light curves depend on the gas density distribution and the opacity of the dust located in the gaseous structures. Because of this, we present color maps of $\rho_{g}$ and $\kappa$ for an azimuthal cut at different phases showing the case with $s_{max,base}=1$mm in Figure~\ref{fig:dens} and \ref{fig:opa}. In Figure~\ref{fig:dens} is clear that the gas follows a stream along the magnetic lines in the dipolar configuration. Note that at different phases, the shape of the stream is different and the modulation by $B_{pol}$ (see equation~\ref{eq:rho_mag}) stands out 
in the plot corresponding to $phase=0.5$. In 
Figure~\ref{fig:opa}, the low values of $\kappa$ are associated to the large value for $s_{max,base}$. A small amount of evaporation is present at the
tip of the stream. 

The spatial distribution of dust (characterized by $\zeta_{stream}$ 
and $s_{max}$) is given using the evaporation algorithm 
(see details in Appendix~\ref{sec-appendix})
such that the aim is to find $s_{max,base}$ consistent with 
$\Delta mag = 0.1$. The parameters used in the evaporation model implies that the dust can survive close to the star as can be seen in Figure~\ref{fig:opa}.

\begin{figure}
   \centering
   \includegraphics[angle=-90,width=8cm]{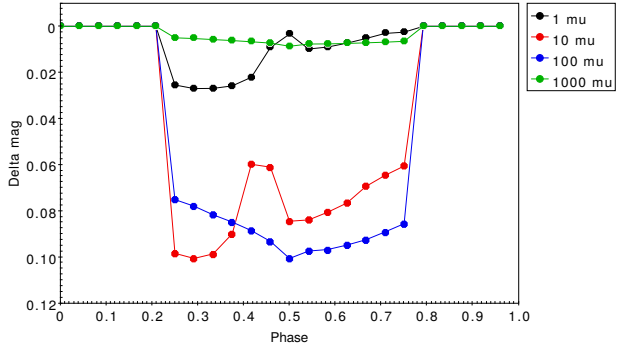}
      \caption{The light curve for V715 Per consistent with 
	       $\dot{M}=5\times 10^{-9}M_{\odot}yr^{-1}$; 
               $s_{max,base}=1\mu$m (black line),
               $s_{max,base}=10\mu$m (red line),
               $s_{max,base}=100\mu$m (blue line),
               $s_{max,base}=1$mm (green line).
              }
         \label{fig:lc}
   \end{figure}

For $s_{max,base}=1$mm, $\Delta mag\sim 0.00724$, more than an order of magnitude lower than the value extracted from the observed light curves. For $s_{max,base}=1\mu$m, $\Delta mag\sim 0.00917$, also less than
expected. Note that these two models, while representing two extreme cases,
produce similar small values for $\Delta mag$. This can be explained because for
$s_{max,base}=1\mu$m, $\zeta_{stream}$ decrease due to strong evaporation and
for $s_{max,base}=1$mm, is not important the evaporation but the opacity of
large grains is small. This lead us to argue that there is a value of 
$s_{max,base}$ in between the previous values able to explain the observed
$\Delta mag$. As shown in the modeled light curves presented in Figure~\ref{fig:lc}, either $s_{max,base}=10$ or $100\mu$m are able to fit the requirement. 
In order to follow the trend in the light curves in Figure~\ref{fig:lc}, we present the models for $s_{max,base}=1,10,100\mu m$, and $1$mm.  
The trend shown in Figure~\ref{fig:lc} is not monotonous, namely 
$\Delta mag$ does not either increase or decrease when $s_{max,base}$ increase. We discuss the various modeled light curves in the following.

The shape of the light curves for $s_{max,base}=1000\mu$m (model A) and 
$100\mu$m (model B) are similar.
The model B has larger $\Delta mag$ than model A which means that 
$\tau_{100}>\tau_{1000}$. Because the mean grain size $<s>$ is smaller in
model B, the opacity $\kappa_{100}$ is larger ($\tau_{100}$ is larger) compared to the value in model A, such that this effect dominates over the decrement of $\kappa_{100}$ due to the evaporation process. 

The shape of the light curves corresponding to $s_{max,base}=100\mu$m (model B) and $10\mu$m (model C) mainly differ because the maximum in $\Delta mag$ shift from the $phase=0.5$ to $phase=0.3$. Model B is in the 
non-evaporating stage and Model C is in the evaporating stage because 
$P_{sat}>>P_{g}$.
The behavior of Model B is completely determined by non-evaporation but in the model C, the evaporation reduces $\zeta_{stream}$ which compensates the increment of $\kappa$ due to the decrement of $<s>$, resulting in $\Delta mag_{100}\sim \Delta mag_{10}$.

The shape of the light curves for $s_{max,base}=10\mu$m (model C) and $1\mu$m 
(model D) are similar. However, $\Delta mag_{1}<< \Delta mag_{10}$, which means
that the strong evaporation increasing from model C to model D leads to a dominant $\zeta_{stream}$ decrement over the $\kappa$ increment associated to a decrement in $<s>$. 

\begin{figure} 
   \centering
   \includegraphics[angle=-90,width=8cm]{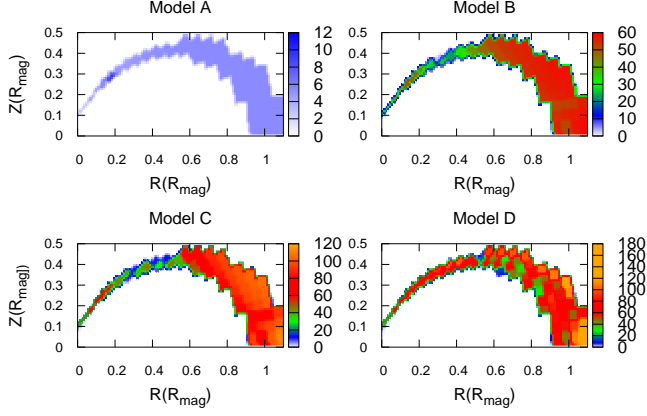}
      \caption{The color map representing the opacity for the $phase=0.3$.  
               This is part of the data used to find the light curves in 
               Figure~\ref{fig:lc}. Here, we show the model A
               ($s_{max,base}=1$mm,upper left plot), the model B
               ($s_{max,base}=100\mu$m,upper right plot), the model C
               ($s_{max,base}=10\mu$m,lower left plot), and the model D
               ($s_{max,base}=1\mu$m,lower right plot).
              }
         \label{fig:opa_phi0.3}
   \end{figure}

\begin{figure} 
   \centering
   \includegraphics[angle=-90,width=8cm]{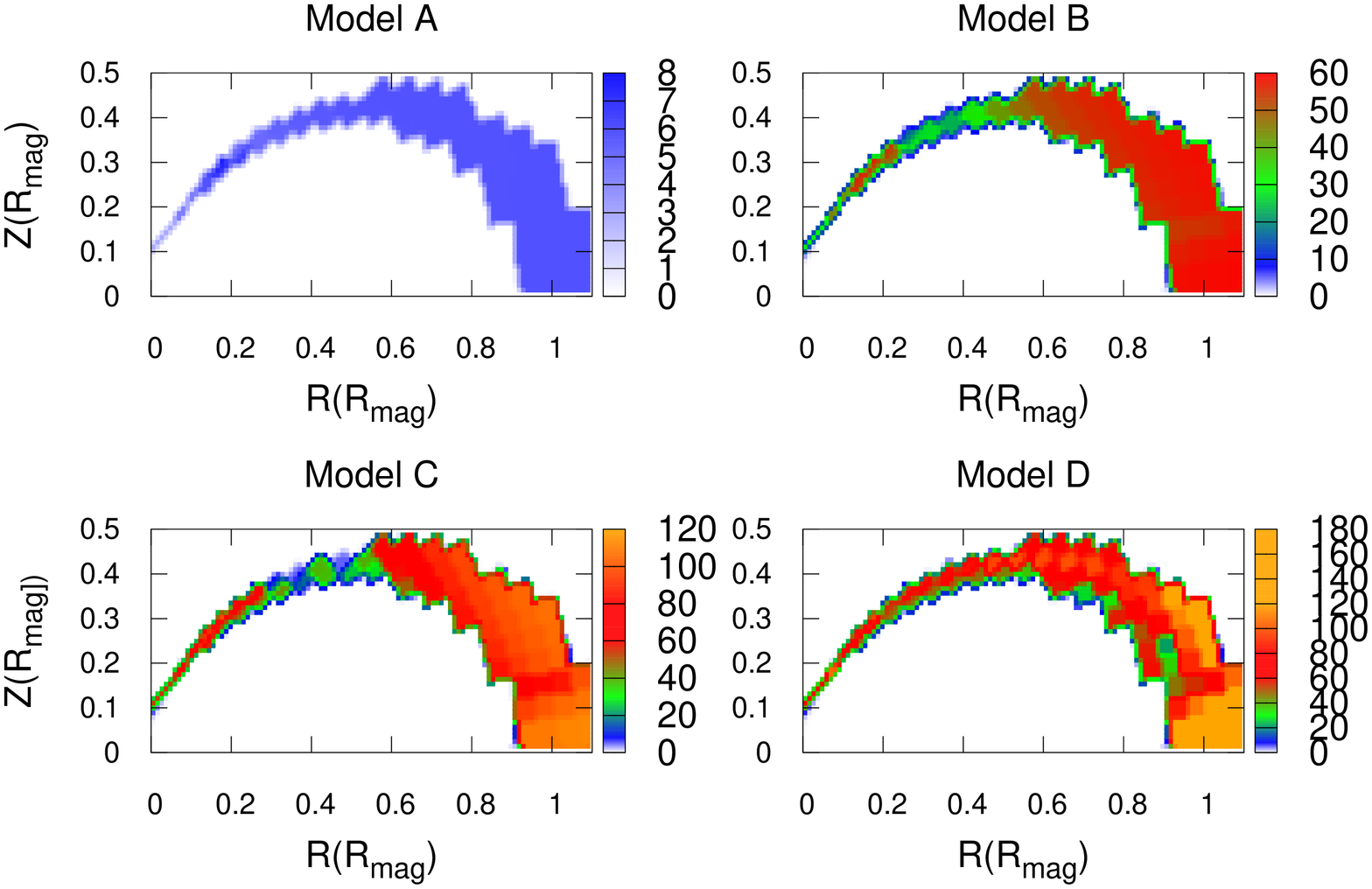}
      \caption{The color map representing the opacity for the $phase=0.5$. 
               This is part of the data used to find the light curves in 
               Figure~\ref{fig:lc}. Here, we show the model A
               ($s_{max,base}=1$mm,upper left plot), the model B
               ($s_{max,base}=100\mu$m,upper right plot), the model C
               ($s_{max,base}=10\mu$m,lower left plot), and the model D
               ($s_{max,base}=1\mu$m,lower right plot).
              }
         \label{fig:opa_phi0.5}
   \end{figure}

The light curves of models B and C are different (see Figure~\ref{fig:lc}).
The main difference is that the maximum of $\Delta mag$ changes from $phase=0.5$
(the azimuthal configuration where the maximum in density is located) in the
model B to $phase=0.3$ in model C. In order to explain how the shape of the light curves changes for different $s_{max,base}$, we present color maps for these two phases. The Figures~\ref{fig:opa_phi0.3} and 
\ref{fig:opa_phi0.5} show $\kappa$ color maps for different values of 
$s_{max,base}$ at $phase=0.3$ and $phase=0.5$, respectively. Note that in the model C, $<s>$ decreases slower in a denser region because $\dot{s}\propto P_{sat}-P_{g}$, but at the same time $\tau$ decreases faster due to the amount of material evaporated. 
Besides $s_{max,base}$ decreases from model B to C, increasing $\kappa$. The result is that in model C at $phase=0.5$ (largest density), the evaporation dominates over the other mechanisms increasing $\tau$ and at $phase=0.3$, the increment of $\kappa$ due to a smaller $s_{max,base}$ dominates over evaporation in a lower density region. This result in a shift of $\Delta mag$ maximum between
models B and C. 

Note that, in the framework of this model, any way the dust is distributed, the only way to explain $\Delta mag = 0.1$ is with an optically thin scenario; a direct estimate of the required mean optical depth is $<\tau>= 0.092$, which is consistent with the best-model presented.

\subsection{The low-$\dot{M}$ state: higher temperature models}
\label{sec-low-state-highT}

We name the models presented in this section, the high temperature models because we assume that the temperature can increase due to heating caused by
the deposition of heat by Alfven waves as mentioned in \citet{hartmann}. As 
explained in Section~\ref{sec-dust-mag}, the dust cannot survive at the high temperatures (up to $\sim 7000$K) present in most of the region limited by 
$R_{mag}$. 

\begin{figure} 
   \centering
   \includegraphics[angle=-90,width=8cm]{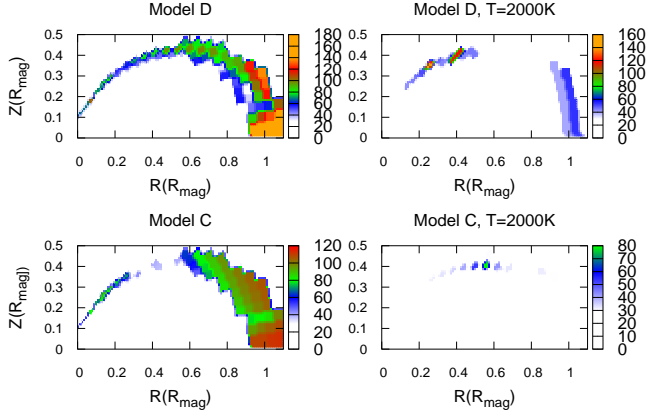}
      \caption{The color map representing the opacity for the $phase=0.5$. 
               Here, we show the fiducial models:
               $s_{max,base}=1\mu$m (upper left plot), and
               $s_{max,base}=10\mu$m (lower left plot) complemented with
	       the high temperature models:
               $s_{max,base}=1\mu$m,$T_{dust}=2000$K (upper right plot), and 
               $s_{max,base}=10\mu$m,$T_{dust}=2000$K (lower right plot).
              }
         \label{fig:opa_smax1_10mu_T2000}
   \end{figure}

\begin{figure} 
   \centering
   \includegraphics[angle=-90,width=8cm]{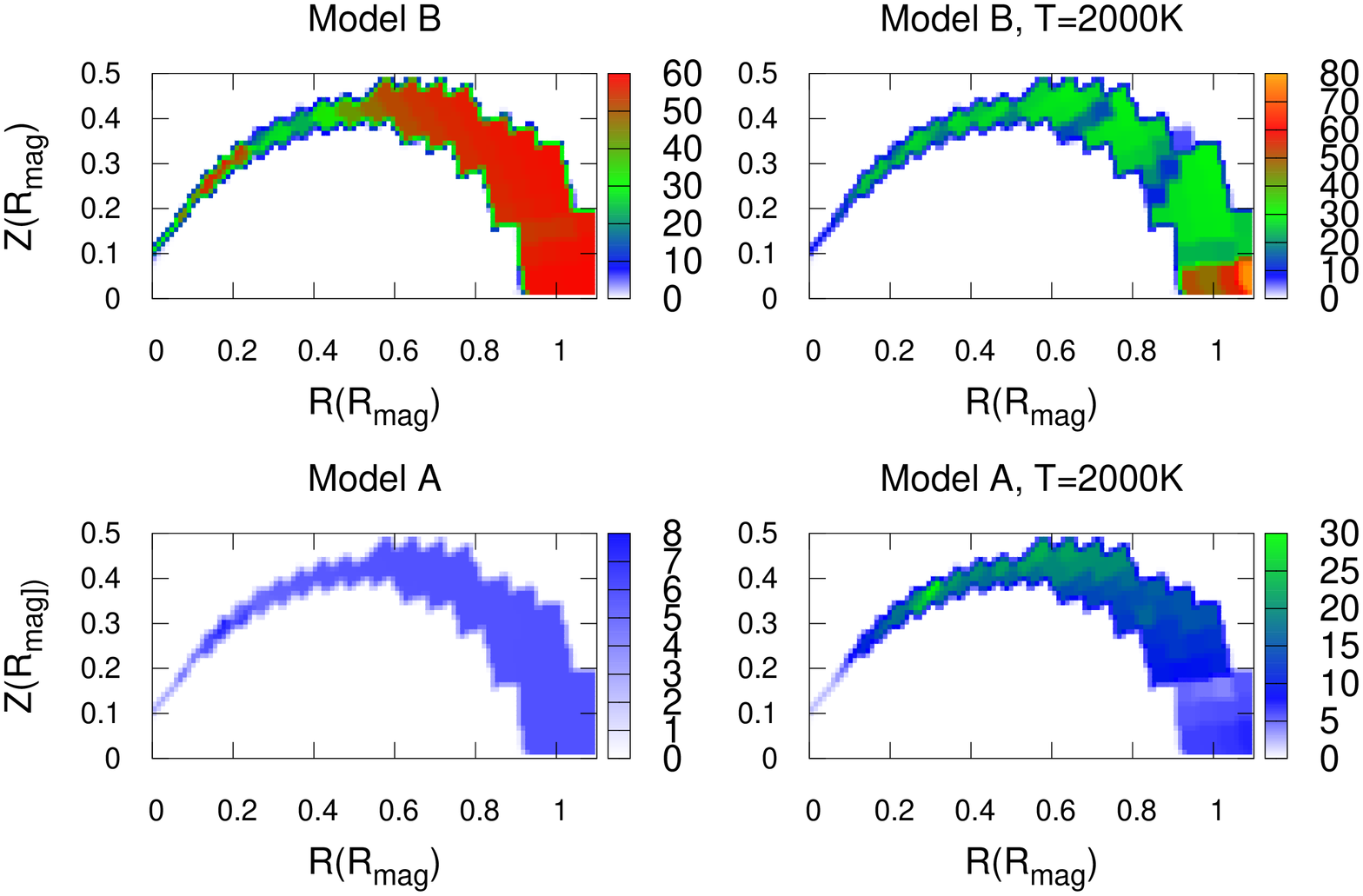}
      \caption{The color map representing the opacity for the $phase=0.5$. 
               Here, we show the fiducial models:
               $s_{max,base}=100\mu$m (upper left plot), and
               $s_{max,base}=1$mm (lower left plot) complemented with
	       the high temperature models:
               $s_{max,base}=100\mu$m,$T_{dust}=2000$K (upper right plot), and 
               $s_{max,base}=1$mm,$T_{dust}=2000$K (lower right plot).
              }
         \label{fig:opa_smax100_1000mu_T2000}
   \end{figure}

In order to analyze the effect of increasing the temperature, we show the opacity
distribution for the cases $s_{max,base}=1,10,100\mu$m and $1$mm when fixing $T_{dust}$ to $2000$K. The color maps are shown in 
Figures~\ref{fig:opa_smax1_10mu_T2000} and \ref{fig:opa_smax100_1000mu_T2000}.

For all the models is clearly seen that there is more evaporation than in
the fiducial models. 
In Figure~\ref{fig:opa_smax100_1000mu_T2000}, we can see that the high temperature model with $s_{max,base}=1$mm shows an increase of the opacity compared with the fiducial model. In other words, $P_{sat}$ increases such 
that $P_{sat}>P_{g}$, in this way reaching the evaporation stage. At this 
stage, the grain size decreases and the opacity increases.
Because of the high inclination, the dust occulting the 
star is located close to $R_{mag}$ where the opacity is the lowest, leading to a low-amplitude light curve which is not able to explain the observed light
curve. For the other models, the opacity decreases, such that these models are not able to explain the observed light curve. Our conclusion is that the survival dust is restricted to a thin layer where the precise width depends on a detailed modeling of the stream heating by Alfven waves.

Note that the high temperature model with $s_{max,base}=1$mm in Figure~\ref{fig:opa_smax100_1000mu_T2000} clearly shows that as the material moves towards the star, the opacity increases due to the decrease of $s_{max}$ but finally there is an opacity decrease because the amount of material evaporated dominates over the decrement of $s_{max}$.

\section{Conclusions}
\label{sec-conclusions}

The main conclusions of the analysis and modeling of the dipper light curve of V715 Per are the following:

1) The dust can survive at $R_{mag}$ where the disk
is truncated for reasonable parameters for the
density of the optically thick disk.

2) Assuming that the gas is distributed following the stellar B-lines 
and using the values for the observed $\dot{M}$, and a standard dust 
to gas ratio without evaporation, the dust in the magnetosphere is optically thick 
($<\tau>=2.76$). However, in order to explain the observations we include
evaporation, such that the required value $<\tau>={2.3\over 2.5}\times\Delta mag={2.3\over 2.5}\times 0.1=0.092$ in the optically thin regime is consistent with the best model.

3) The low gas density and the radial range
of the magnetosphere results that in most of it
$T_{dust}>T_{sub}$ , such that the dust cannot survive.
Yet, the dust is able to survive in a shell with an
outer radius given by $R_{mag}$ , because the timescales
for evaporation ($t_{sub}$) and free-fall towards the
stellar surface ($t_{ff}$ ) are similar.

4) We find a model using as a free parameter, the largest size
of the grains at the base of the magnetospheric stream, $s_{max,base}$.
We find that the models with $s_{max,base}=10,100\mu$m are able to explain the
observed amplitude of the optical light curve ($\Delta mag$). 
The grains move towards the star and evaporate, changing their size and shaping the light curve. 

5) The analysis of the models presented in Section~\ref{sec-low-state} suggests that two competing mechanisms  are simultaneously responsible for the shape of the the light curves. The first one is the evaporation that reduces the dust density leading to a decrement of the dip amplitude. However, during this process, the mean size of the grain decreases and the dip amplitude increases because the grain opacity increases. 
As $s_{max,base}$ changes, the contribution of both mechanisms also changes, thus the depth and shape of the light curves vary accordingly.

6) In the modeling, we do not include a model of deposition of heat by Alfven waves in the stream as suggested by \citet{hartmann}, which would considerably increase the temperature inside the magnetosphere. At such a high temperature ($\sim$7000~K) the grains evaporate almost instantly, thus restricting the dust to a thin layer very close to $R_{mag}$. In any case, due to the high inclination of the system, almost all the blocking structure is close to 
$R_{mag}$, a fact that can be checked in any of the color maps where the shape of the stream can be seen.

7) Because the values for $M_{\star}$ (the main parameter
to calculate $t_{ff}$) and $\dot{M}$ (the main parameter to
calculate $R_{mag}$ ) for V715 Per are typical of many
other systems showing this behavior in their light curves, this mechanism 
can be invoked to explain the light curves 
where the flux shows color changes associated to circumstellar extinction, which indicates the occulting dust is optically thin.

\begin{acknowledgements}
This project has received funding from the European Research Council (ERC) under the European Union's Horizon 2020 research and innovation program (grant agreement No 742095 ; SPIDI : Star-Planets-Inner Disk- Interactions; http://www.spidi-eu.org).
E.N. appreciate the support at the Institut de Planétologie \& d’Astrophysique de Grenoble (IPAG) by Observatoire des Sciences de l'Univers de Grenoble/Université Grenoble Alpes (OSUG/UGA) while in a sabbatical stay where part of this work was done.  
\end{acknowledgements}

\begin{appendix} 
\section{Abundance, opacity and grain size distribution of the dust in
the evaporating magnetospheric stream.}
\label{sec-appendix}

As the grains in the magnetospheric stream moves towards the star, $P_{sat}$ 
surpass $P_{gas}$ and the dust starts to evaporate. This process is not instantaneous, the grain decreases its size at a rate given by $\dot{s}$ as
it is described in Section~\ref{sec-dust-mag}. During this process, all the
grains decrease in size, in the following we will show that the power-law
grain size distribution is conserved.

The number of grains within the size range [s,s+ds] are: $Cs^{-3.5}ds$, 
with $C$ a normalization coefficient. The number of grains
is conserved after the span of time dt, if $ds$ changes to $\dot{s}dt+ds$. 
The size distribution, $f(s,t)$, satisfies the equation

\begin{equation}
 f(s,t)\times (\dot{s}dt+ds)=Cs^{-3.5}ds,
\end{equation}

which can be written as

\begin{equation}  
 (Cs^{-3.5}-f(s,t))ds-f(s,t)\dot{s}dt=dg(s,t)=0.
\end{equation}

Here $g$ is a constant. From the total differential we can identify
the terms $\partial g\over \partial s$ and $\partial g\over \partial t$ 
and using the fact that  
${\partial^2 g\over \partial s\partial t}={\partial^2 g\over \partial 
t\partial s}$,
we found that  

\begin{equation}
 -{\partial f\over \partial t}=\dot{s}{\partial f\over \partial s}.
\end{equation}

The previous equation can be solved using the initial condition 
$f(s,t=0)=Cs^{-3.5}$ such that $f(s,t)=Cs^{\prime -3.5}$, where
$s^\prime = s-\dot{s}t$ is the size of the grain after time $t$. Note
that the power-law is conserved with the same exponent.

In a stationary configuration, each location of the stream has assigned a value
for $s_{max}$. In the ISM and in the disk, the typical size distributions are
characterized with a $s_{min}=0.005\mu$m, for simplicity we keep this value here.
For the estimate of $s_{max}$ we use $\dot{s}$ which requires
the radial velocity given in \citet{hartmann} calculated for ballistic infall from rest at $R_{mag}$ used to calculate the density, such that $s_{max}$ decreases as the dust moves closer to the star.

The dust opacity at each point is calculated as a time dependent process starting
with the opacity of the large grains distribution ($s_{max}=s_{max,base}$) associated 
with the dust in the disk finally arriving to the case where $s_{max}\geq 0.25\mu$m,
$s_{max}=0.25\mu$m is the value associated to a small grains distribution typical of the ISM, the lowest value considered here. A grain
with $s_{max} < 0.25\mu$m completely evaporates in a span of time small compared
to the time it takes to arrive to this point. Thus, we assume that there is
dust in a streamline if $s_{max}\ge 0.25\mu$m.

The opacity associated to the dust distributed with $0.25\mu m<s_{max}<1mm$
is calculated interpolating between the opacity of the large grains and the 
small grains distribution of the standard abundance, $\zeta_{std}=0.0065$. 

We correct the opacity including the fact that the dust abundance decreases 
as each grain decreases in radius in the gradual evaporation of its superficial
mass. This process should be done starting from the base of the stream.

The correction factor is calculated using the total initial dust mass 
($M_{init}$) in a volume $\Delta V$ and the mass evaporated $M_{evap}$ in a time span $\Delta t$.

From the initial dust abundance $\zeta_{init}$, 
$M_{init}=\zeta_{init}\rho_{g}\Delta V$, where $\rho_{g}$ is the gas density.
On the other hand, we can write

\begin{equation}
 M_{init}=\int^{s_{max}}_{s_{min}}Cs^{-3.5}\rho_{dust}{4\pi\over 3}s^{3}ds.
\end{equation} 

Equating both expressions

\begin{equation}
 C={3\zeta_{init}\rho_{g}\Delta V\over 8\pi\rho_{dust}[s_{max}^{0.5}-s_{min}^{0.5}]}.
\end{equation}

After the integration time $\Delta t$, the evaporated dust mass is

\begin{equation}
 M_{evap}=\int^{s_{max}}_{s_{min}}Cs^{-3.5}\rho_{dust}{4\pi\over 3}(s^3-
 (s-\dot{s}\Delta t)^3)ds.
\end{equation}

The corrected dust abundance $\zeta$ is calculated using

\begin{equation}
 \zeta\rho_{g}\Delta V=M_{init}-M_{evap},
\end{equation}

and doing all the straightforward integrations.

Now, for the next integration time $s_{max}=s_{max}-\dot{s}\Delta t$ and
$\zeta_{init}=\zeta$ and this algorithm is repeated until the requested 
stream point is reached.

\end{appendix}
\end{document}